\documentclass[%
 reprint,
 amsmath,amssymb,
 aps,
]{revtex4-2}
\usepackage{color}
\usepackage{natbib}
\usepackage{indentfirst}
\usepackage{graphicx}
\usepackage{dcolumn}
\usepackage{bm}

\begin{document}
\title{Sound attenuation derived from quenched disorder in solids}

\author{Bingyu Cui$^{1}$, Alessio Zaccone$^{1,2,3}$, Eugene Terentjev$^{1}$}
\email{emt1000@cam.ac.uk}
\affiliation{${}^1$Cavendish Laboratory, University of Cambridge, JJ Thomson Avenue, CB3 0HE Cambridge, U.K.}
\affiliation{${}^2$Department of Chemical Engineering and Biotechnology, University of Cambridge, Philippa Fawcett Drive, CB3 0AS Cambridge, U.K.}
\affiliation{${}^3$Department of Physics ``A. Pontremoli", University of Milan, via Celoria 16, 20133 Milano, Italy}

\begin{abstract}
\noindent In scattering experiments, the dynamical structure factor (DSF) characterizes inter-particle correlations and their time evolution. We analytically evaluated the DSF of disordered solids with disorder in the spring constant, by averaging over quenched disorder in the values of lattice bond strength, along the acoustic branch. The width of the resulting acoustic excitation peak is treated as the effective damping constant $\Gamma(q)$, which we found to grow linearly with exchanged momentum $q$. This is verified by numerically calculating a model system consisting of harmonic linear chains with disorder in spring constant. We also found that the quenched averaging of the vibrational density of states produces a characteristic peak at a frequency related to the average acoustic resonance. Such a peak (the excess over Debye law) may be related to the ``boson peak'' frequently discussed in disordered solids, in our case explicitly arising from the quenched disorder in the distribution of spring constants.
\end{abstract}

\pacs{}
\maketitle

\section{Introduction}
Disordered solids exhibit many anomalous features compared with their crystalline counterparts. Specifically, the mechanism of sound attenuation in non-perfect crystals and disordered materials remains elusive \cite{Phillips2015,Frick1995,Angell1995}. Many features of elastic wave propagation are stored in the dynamical structure factor (DSF), which is directly measured in X-ray or inelastic neutron scattering experiments, or can be calculated in numerical simulations. In particular, the width of the structural peaks in DSF is representative of the mechanical damping in the material, which is a function of the wave vector: $\Gamma(q)$. 
Interaction of sound waves and thermal phonons in an isotropic or anisotropic elastic solid was systematically studied \cite{Herring1954,Zyryanov1966}, and the quadratic scaling
$\Gamma(q) \sim q^2$ was found in the low frequency region. Such a quadratic scaling of damping constant in both longitudinal and transverse acoustic waves is known as the Akhiezer damping (1938) \cite{Akhiezer1938,Herring1954}. On the other hand, a linear scaling of the damping coefficient, $\Gamma \sim q$, was derived by Landau and Rumer in 1937 \cite{Landau1937,Herring1954,Hasson1975}, considering crystals with anharmonicity in their bond potential, and obtaining the damping coefficient for transverse waves. The Landau-Rumer calculation was carried out when quantum effects are dominant, so that $\Omega\tau \gg 1$, where $\Omega$ and $\tau$ are the wave propagation frequency and relaxation time of the thermal phonons, respectively \cite{Zyryanov1966}. 

In an attempt to elucidate the characteristics of sound waves, theoretical investigations and also numerical simulations accounting for the normal modes in both the harmonic approximation and anharmonic theory have been used. 
I. M.  Lifshitz discovered that a single defect produces a narrow  peak  in  the  VDOS~\cite{Lifshitz_defect1,Lifshitz_defect2}. The fact that the width of the peak is finite  suggests  that  the  vibration  is  not  really  local,  and ever since these vibrational states are called ``quasi-localized vibrations". Those ideas were first put in practice in the context of the so-called boson peak in the VDOS and the closely-related low-T thermal anomalies of glasses~\cite{Klinger}.

Schirmacher \textit{et al.} constructed a cubic lattice of coupled classical harmonic oscillators with spatially fluctuating nearest neighbor force constants, and also found a low-frequency excess peak in the scaled density of states $D(\omega)$ \cite{Schirmacher1998}. Accounting for the effect of spatial
wavelength fluctuations, Montagna \textit{et al.} presented a mechanism reproducing the $\sim q^2$ dependence of the broadening of the Brillouin peaks at high frequency \cite{Montagna1999}. At low frequency, the $q^2$ law can be explained in terms of momentum diffusion due to elastic disorder-induced scattering~\cite{Baggioli}. 

Nevertheless, theory has somehow been left behind. The main theoretical methodology to describe sound wave propagation in disordered materials is based on the replica-trick for heterogeneous elasticity, which amounts to self-consistent Born approximation of the propagating wave and its self-energy \cite{Cui2019soft,Maurer2004,Schirmacher2006}. This approach predicts a Rayleigh scattering $\sim k^{d+1}$ or $\sim \omega^{d+1}$, for sound attenuation in an intermediate range of $k$, or $\omega$. These predictions are based on assuming a Gaussian fluctuation of elastic moduli throughout the material.
In simulations and experiments of sound attenuation in glasses, however, the sound attenuation is not of the Rayleigh type, but rather has an important logarithmic correction, $\sim (-k^{d+1} \ln k)$ \cite{Kinder1979,Giordano2009,Baldi2010,Baldi2011,Ruta2012,Moriel2019,Flenner,Mizuno2018}. The first successful theoretical prediction of this generalized Rayleigh scattering law for glasses has been achieved in Ref.\cite{Cui2020}, by implementing anistropic long-ranged power-law correlations of elasticity, which is supported by recent experimental and theoretical works~\cite{Fuchs2017,Wang2020}.

These important recent advances are, however, focused on an intermediate range of $k$, and there is still no analytical approach so far to address the sound damping mechanism in the practically relevant range of low wavevectors. We will make a step forward in studying the DSF and the associated damping effects in a lattice with quenched random disorder. The quenched (or frozen-in) disorder occurs when one mixes some random impurities in a melt and then cool it down rapidly so that the host and impurities are not in thermal equilibrium. Rather, the impurities remain blocked in random fixed positions. Quenched disorder is static, and leads to new phenomena and ideas not only in spin glasses, but also in many other glassy systems \cite{Edwards1975, Benetatos2010}.

In this article, we account for the disorder via a Gaussian distribution of spring constants in a perfectly ordered lattice, the Gaussian variance reflecting the degree of disorder. The sound (elastic) wave is replaced by the lattice wave considered in harmonic theory, as the two are equivalent for low wavevectors $q$. In this way we obtain the observable quenched-averaged DSF, which acquires a characteristic broadening of structural peaks that allows us to derive the analytical expression for the damping coefficient $\Gamma (q)$. We find $\Gamma$ is proportional to the frequency of the acoustic phonon mode, and therefore is linear in $q$ at low wavevectors, a result that turns out to be valid in all dimensions. This analytical result is confirmed by a numerical simulation of elastic lattice where we carry out the explicit averaging over the quenched ensemble of random systems. This allows an estimate of the effective disorder strength for realistic materials where the damping coefficient $\Gamma$ is independently measured.
We also found that the quenched averaging of the vibrational density of states (vDOS) produces a characteristik peak at a frequency related to the average acoustic resonance. Such a peak (in excess over the Debye law) may be interpreted as the ``boson peak'' frequently discussed in disordered solids, in our case explicitly arising from the quenched disorder in the distribution of spring constants.

\section{Dynamical structure factor}

In a system of $N$ particles located at lattice positions $\mathbf{r}_I,I=1,2,...,N$, the DSF is defined from the density-density correlated function, via both spatial and time Fourier transformation:
\begin{equation}
S(\mathbf{q},\omega)=\frac{1}{N}\int_{-\infty}^{\infty}\sum_{I,J}\left\langle e^{(i\mathbf{q}\cdot 
\mathbf{r}_{J}(t)-i\mathbf{q}\cdot \mathbf{r}_I(0))}\right\rangle e^{-i\omega t}\frac{dt}{2\pi},
\end{equation}
where the symbol $\langle.\rangle$ refers to the thermal average. In a lattice, the position of the particle $I$ in cell $l$ can be written as $\mathbf{r}_I=\mathbf{R}_I+\mathbf{u}_I$,
where $\mathbf{R}_I$ and $\mathbf{u}_I$ are the equilibrium position and displacement, respectively. This applies to a general cell $l$, and so we suppress this index in the following. In the harmonic bond approximation, the longitudinal DSF of a one-phonon process in classical limit is the sum over normal modes \cite{Ashcroft1976,Shintani2008}:
\begin{align}
S(\mathbf{q},\omega)=&
\frac{k_BT}{2n_c}\sum_s\left|\sum_{I}\frac{\mathbf{q}\cdot\mathbf{e}_I^s(\mathbf{q})}{\omega_s(\mathbf{q})\sqrt{M_I}}e^{i\mathbf{q\cdot}
\mathbf{R}_I}\right|^2\notag\\
&\quad \  \times \left(\delta(\omega-\omega_s(\mathbf{q}))+\delta(\omega+\omega_s(\mathbf{q}))\right).
\label{eq:dsfscattering}
\end{align}
Here, $s=1,2,...,(Nd)$ labels the normal modes, $\mathbf{e}^s_I(\mathbf{q})$ and $\omega_s(\mathbf{q})$ are the eigenvector and eigenvalue of the dynamical matrix elements, corresponding to the $I$th particle (mass $M_I$), and  $n_c$ is the number of particles within the unit cell.  In the classical limit, the dynamical structure factor and the Green's function of the elastic wave equation are related by the fluctuation-dissipation theorem (FDT) \cite{Kubo1957,Maurer2004,Schirmacher2015a,Schirmacher2015b}:
\begin{align}
S(\mathbf{q},\omega)= & \frac{k_BTq^2}{\pi\omega}\text{Im}[G(\mathbf{q},\omega)]
\label{eq:FDT} \\
\mathrm{with} & \quad  G(\mathbf{q},\omega)=\frac{1}{-\omega^2+q^2v^2(\omega)} , \notag
\end{align}
where $v(\omega)$ is the (generalized, complex) sound velocity related to the frequency-dependent linear elastic modulus $C(\omega)=\rho v^2(\omega)=C'(\omega) - i C''(\omega)$. The form of the Green's function in Eq. \eqref{eq:FDT}  represents elastic waves propagating in the system where elasticity presents spatial correlations. In the standard way, defining the sound attenuation (damping) coefficient $\Gamma(\omega)=\omega C''(\omega)/C'(\omega)$, and the resonance frequency $\Omega(q)=q\sqrt{C'(\omega)/\rho}$, it can be directly shown that the DSF becomes \cite{Schirmacher2015b}:
\begin{equation}
S(q,\omega)=\frac{k_BTq^2}{M\pi\omega}
\frac{\Omega^2(q)\Gamma(\omega)/\omega}{[\Omega(q)^2-\omega^2]^2+[\Omega^2(q)\Gamma(\omega)/\omega]^2}.
\label{eq:4}
\end{equation}
Near the resonance frequency $\omega\sim\Omega(q)$, the form of DSF can be approximated in the form of damped harmonic oscillator (DHO):
\begin{equation}
S(q,\omega) \approx \frac{k_BTq^2}{M\omega\pi}\frac{\omega\Gamma(q)}{(\Omega^2(q)-\omega^2)^2+\omega^2\Gamma(q)^2}.
\label{eq:DHO}
\end{equation}
Expression \eqref{eq:DHO}, for a fixed $q$, reaches a maximum value (peak) when $\omega_\mathrm{max}^2 =\Omega^2-\Gamma^2/2$. Assuming positive parameters $\Gamma$ and $\Omega$, the half width at half maximum (HWHM) takes the form
\begin{align}
\left(\omega^2-\omega_\mathrm{max}^2\right)=\Gamma\sqrt{\Omega^2-\Gamma^2/4}.
\label{eq:DHOHWHM}
\end{align}
Thus, near the resonance frequency $\Omega$ and at low damping, $\Gamma$ is roughly the width of the broadened dynamic structure factor peak. We also note that, at $\Gamma\rightarrow 0$, $S(q,\omega)$ reduces to the sum of $\delta$-functions:
\begin{equation}
\lim_{\Gamma\rightarrow0}S(q,\omega)=\frac{k_BTq^2}{2M\omega^2}[\delta(\omega-\Omega)+\delta(\omega+\Omega)],
\label{eq:dsfdelta}
\end{equation}
where the identity $2a\delta(x^2-a^2)=\delta(x+a)+\delta(x-a)$ was used. Expression \eqref{eq:dsfdelta} in the $\Gamma\rightarrow 0$ limit reflects the perfect crystal limit. To depict this, in Appendix, we provide plots of DSF for 1D linear chains connected with spring having some variation in spring constants (representing a realization of disorder), which already show broadening of $\delta$-function peaks, indicating that emergence of sound attenuation (damping) is linked to disorder in perfect harmonic crystals.

\section{Quenched random disorder}


When the local elastic modulus, or the spring constant $k$ of a harmonic bond in a discrete lattice model, takes random values in the system, all quantities like DSF should be averaged over the distruibution of such random disorder. In general, quenched disorder could occur in three ways: by ramdomizing masses, spring constants, or equilibrium lattice positions. It is commonly accepted in scattering theories that first two types of disorder are essentially equivalent \cite{Gelin2016}, so we let all particles have the same mass $M$. We note that disorder in the initial equilibrium positions might also cause the breaking of inversion symmetry, thus changing the affine force fields \citep{Lemaitre2006,Milkus2016,Cui2020}, which we are not considering here. 

We consider the random distribution of values of the spring constant, assuming they satisfy a normal distribution, $k_i\sim \mathcal{P}(k_i)\equiv N(k_0,\sigma)$, where $i$ is the index of the $i$-th spring,  the mean value $k_0$ is the measure of material stiffness, and the variance $\sigma$ serves as the measure of disorder. In principle, the spring constant should always be non-negative, so one can introduce the truncated Gaussian distribution to set a lower bound for spring constant \cite{Schirmacher1998}, namely $\mathcal{P}'(k)=\mathcal{P}(k)\Theta(k-k_{min})$ for some $k_{min}>0$. However, we found that the truncation does not change any of our results (analytical and numerical) since we always work in the regime of relatively low disorder: $\sigma/k_0 < 1$. 
Then the quenched averaging of DSF as given in Eq. \eqref{eq:dsfscattering} is easy due to the effect of $\delta$-functions in integration (which is why the truncation of the Gaussian range has no effect in our case),
\begin{align}
&\langle S(\mathbf{q},\omega)\rangle=\int\prod_i\mathcal{D}[k_i]\mathcal{P}(k_i)S(\mathbf{q},\omega) \\
&= \frac{f(q,L)}{\omega} \int\prod_i\mathcal{D}[k_i] e^{\sum_i-\frac{(k_i-k_0)^2}{2\sigma^2}}\delta(\omega^2-\omega^2_{A,L}(\mathbf{q},k_i)),\notag
\end{align}
where $f(q,L)$ represents the $q$ dependent part: the prefactor of $\delta$-functions in Eq. \eqref{eq:dsfscattering} and depends on the characteristic length $L$ (hence the size) of the system since when the systematic size is large enough, sound wave is expected to be localized, known as Anderson localization \cite{Anderson1958,Ziman1979} Since this localization always occur, regardless of type of disorder and excited mode, we decouple it from the elastic disorder and write it implicitly in the form of function $f(q,L)$.
In the second equality we focus on the longitudinal acoustic mode, since only the acoustic branch $\Omega=\omega_{A,L}(q)$ contributes to elastic waves \cite{Born,Cui2020}. That means only one out of $Nd$ modes survives, and is then subjected to quenched averaging. Although this still leaves us one set of $\delta$-functions, which looks similar to Eq. \eqref{eq:dsfdelta}, the physics is completely different. Here, the system with quenched random disorder is averaged over local realizations of the random variable (the spring constant), and we will shortly discover that it has a non-zero damping coefficient, while in Eq. \eqref{eq:dsfdelta} we had a perfect crystal without damping in sound waves. 


We also note that, unlike Eq. \eqref{eq:FDT}, this DSF relates to the lattice waves (or collective phonon modes). In general, the lattice waves and the elastic waves are different \cite{Born,Trinkle2008}, however, their difference is negligible when $q$ is small.  
In principle, the acoustic branch $\omega_{A,L}(q)$ receives contributions from all discrete springs satisfying the Gaussian distribution. For analytical clarity, we simplify this by using a single continuous variable $k$ for the spring constant which is treated as a random variable. The quenched average becomes
\begin{equation}
\langle S(q,\omega) \rangle \propto \frac{f(q,L)}{\omega}\int \mathrm{d}k \  e^{-\frac{(k-k_0)^2}{2{\sigma^2}}}\delta(\omega^2-{\omega}^2_{A,L}(\mathbf{q},k))
\label{eq:avS2}.
\end{equation}
 Performing the integral over $k$ gives 
\begin{equation}
\langle S(q,\omega)\rangle \propto \frac{f(q,L)}{\omega} \exp \left[-\frac{(\omega^2-\omega^2_{A,L}(\mathbf{q},k_0))^2}{2({\sigma}/k_0)^2\omega^4_{A,L}(\mathbf{q},k_0)}\right].
\label{eq:quenchS1}
\end{equation}
The acoustic phonon frequency normally would take the form $\omega^2_{A,L} = 4(k_0/M) \sin^2(qa/2)$, where $a$ is the lattice spacing along the $\mathbf{q}$ direction. We find that, in our averaged disordered system, the expression is changed very little.
The maximum in Eq. \eqref{eq:quenchS1} (the position of the acoustic resonance peak), is found at: 
\begin{equation}
\omega^2_\mathrm{max}=\frac{1}{2} \omega_{A,L}^2(1 - \sqrt{1-2\sigma^2/k_0^2}).  \label{eq:max}
\end{equation}
If we assume low disorder, ${\sigma}^2/k_0^2\ll 1$, then $\omega_\mathrm{max} \approx \omega_{A,L}(\mathbf{q},k_0)$ at the original resonant frequency. Also in this case, the disorder-averaged DSF roughly reaches its half maximum when the exponent in Eq. \eqref{eq:quenchS1} is of order 1. This gives the estimate of the half-width 
\begin{equation}
\omega-\omega_{A,L} \approx \frac{\sqrt{2}}{2}\frac{{\sigma}}{k_0}\omega_{A,L}(\mathbf{q},k_0)
\label{eq:lineardamping}
\end{equation}

\begin{figure}
\centering
\includegraphics[width=0.49\textwidth]{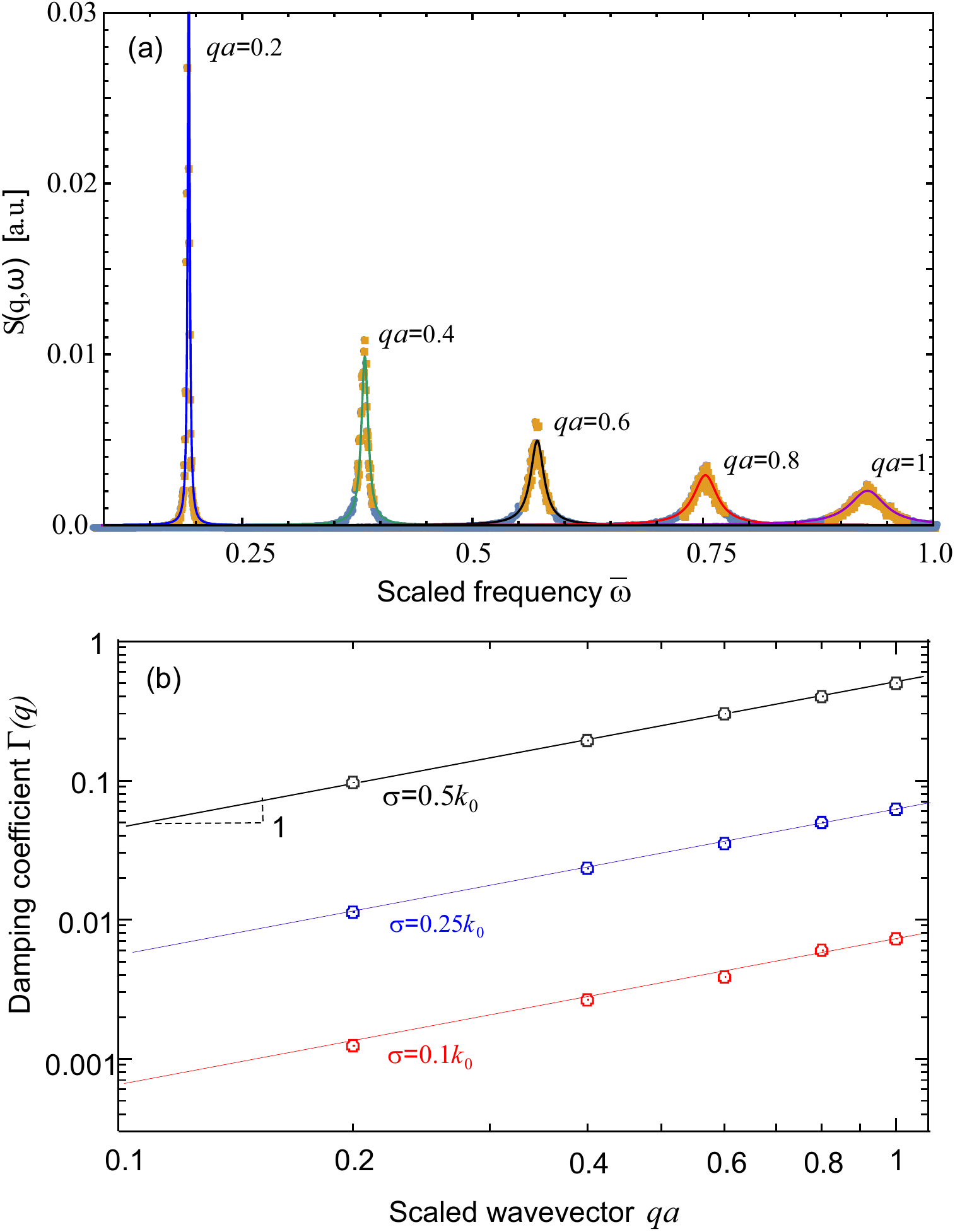}
\caption{(a) The numerical verification of the effect of Gaussian disorder in spring constants on damping, for $\sigma = 0.25 k_0$. There are 20000 masses separated in equal distance in a unit cell. The averaged DSF $\langle S(q,\omega) \rangle$ for several values of $q$, numerically calculated using the Gaussian distribution in spring constants of 1D linear chain, plotted against the non-dimensional frequency scaled by a factor $\sqrt{k_0/M}$. The data for each $q$ is fitted by DHO in Eq. \eqref{eq:DHO} (solid line).  (b) The log-log plot of the damping constant $\Gamma$, obtained as the peak width (also scaled in units of frequency  $\sqrt{k_0/M}$), plotted against the reduced wavenumber $qa$ shows the distinct linear scaling  $\Gamma \sim q$ discussed in the text. }
\label{fig:springs}
\end{figure}

This means that at low disorder the damping constant, (the width of the broadened DSF resonance peak as elucidated by Eq. \eqref{eq:lineardamping}, takes the explicit form: $\Gamma=\sqrt{2}{\sigma}\omega_{A,L}/k_0$. Notably, this scales linearly at low wavevectors, which is the main finding of his work. To verify this observation, we numerically compute the DSF of 1D linear chains, calculating the quenched-averaged DSF for 30 realizations of the random distribution of spring constants between the bonds, and average every point for a given $\omega$, thus having an independent approximate numerical evaluation of $\langle S(q,\omega)\rangle $. This is plotted in Fig. \ref{fig:springs}(a), for an example case of $\sigma = 0.25 k_0$, and several values of $q$ on the same plot, with the numerical data fitted by Eq. \eqref{eq:DHO} to determine $\Gamma(q)$ for each scattering peak. 

Figure \ref{fig:springs}(a) shows the effect of quenched disorder is more pronounced with increasing $q$. This is because the bigger the space we average, the lower $q$ is, and so the less average disorder is left in it. After fitting each peak of the averaged DSF, we obtain the relation between the wavenumber and the effective damping coefficient $\Gamma (q)$ defined in Eq. \eqref{eq:DHO}; this relationship is shown in Fig. \ref{fig:springs}(b) for several values of $\sigma$. The values of damping coefficient differ strongly for the cases of different disorder, but the logarithmic scale of the plot allows us to see the $\Gamma (q)$ scaling for all considered cases. Importantly, we rigorously confirm the linear dependence $\Gamma(q) \sim q$, which is an unexpected result never reported in the literature before. 

\begin{figure}
\centering
\includegraphics[width=0.48\textwidth]{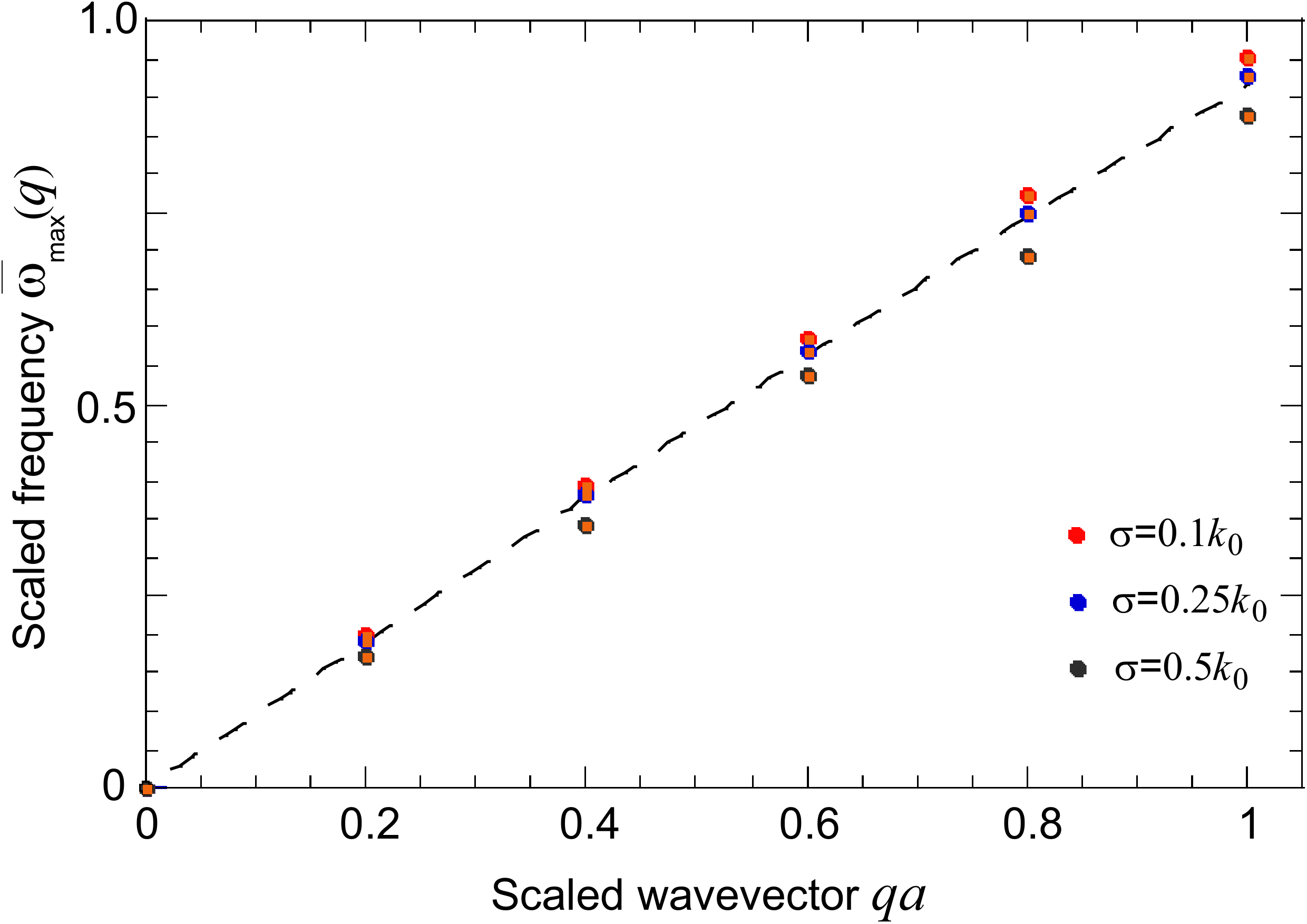}
\caption{The resonant frequency $\bar{\omega}_\mathrm{max}(q)$, scaled by $\sqrt{k_0/M}$, obtained from numerical calculation with Gaussian disorder in spring constants, for several values of $\sigma $, see Fig. \ref{fig:springs}(a). The dashed line is the parameter-free curve $\bar{\omega} = 2 \sin (qa/2)$, which fits the data well, and we can see that deviations from linearity are small, and the dependence on disorder strength $\sigma$ is weak, see Eq. \eqref{eq:max}. }
\label{fig:omega}
\end{figure}


The other fitting parameter for the numerically averaged DSF, illustrated in Fig. \ref{fig:springs}(a), is the resonance frequency $\Omega (q)$ defined in Eq. \eqref{eq:DHO}, which corresponds to our $\omega_\mathrm{max}(q)$ in Eq. \eqref{eq:max}. The comparison of our model with the ideal dispersion relation ${\omega}_{A,L} = \sqrt{4k_0/M} \sin (qa/2)$  supports its validity, especially when the wave vector $q$ is not large. Here we find only a weak dependence on the strength of disorder $\sigma$, see Fig. \ref{fig:omega}. We have to add at this point that the estimated errors in the described procedure of averaging over several disorder realizations are quite low, and we are not showing them in the plots.

\section{Density of states and the characteristic peak}

\begin{figure}
\includegraphics[width=0.48\textwidth]{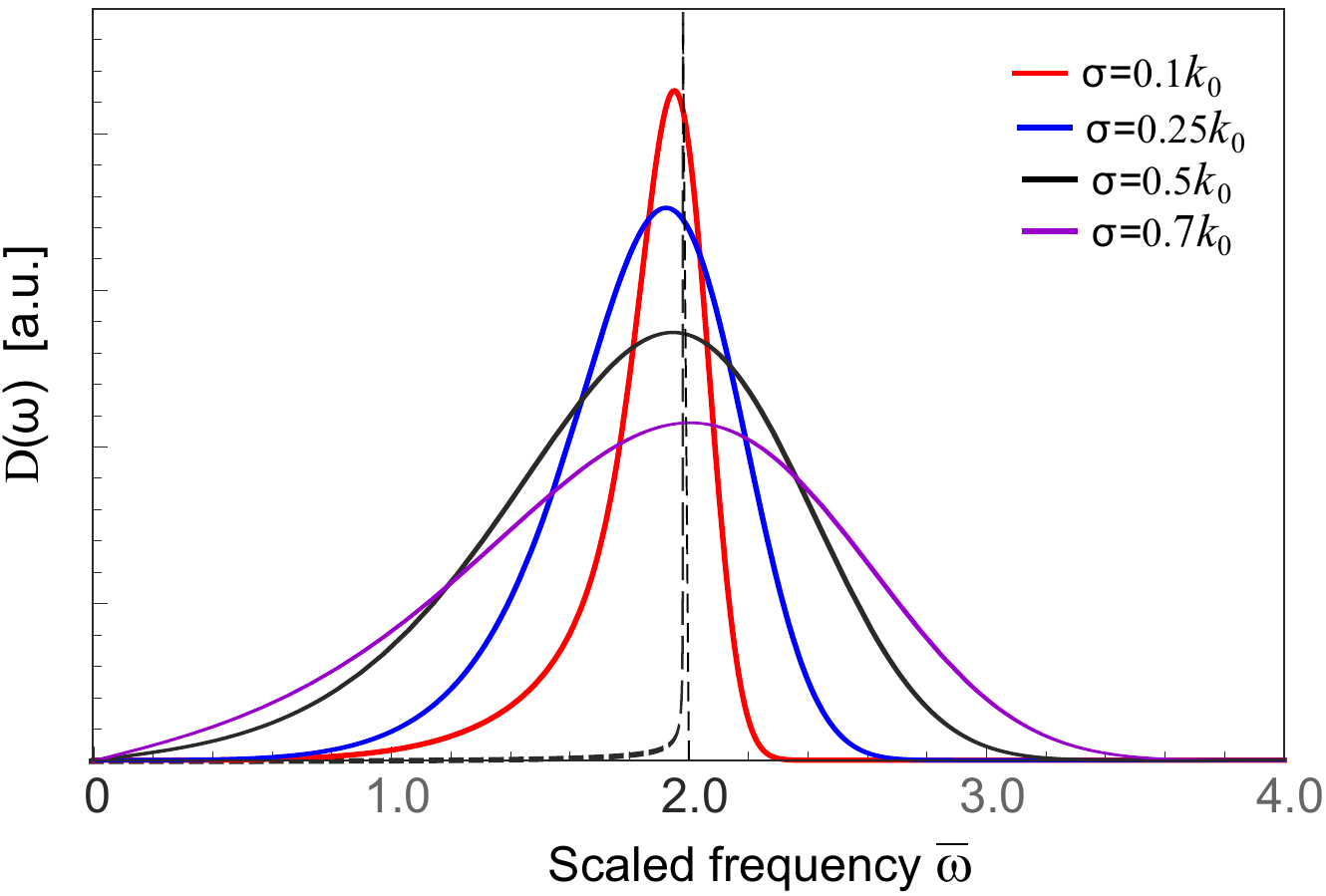}
\caption{Scaled DOS calculated using Eq. \eqref{eq:DOSquench} with several values of ${\sigma}$ listed in the plot. Frequency is scaled in the same way as in Fig. \ref{fig:springs}. The cut-off $q_D$ is chosen at $\pi/a$. In the limit of vanishing disorder, the ``characteristic peak'' transforms to a sharp resonance at a characteristic frequency of $2\sqrt{k_0/M}$ (in that similar to what one obtains from the analysis of an isolated defect~\cite{Lifshitz_defect1,Lifshitz_defect2,Kosevich}), obtained from the integration Eq. \eqref{eq:DOSquench} shown by the dashed line.}
\label{fig:BP}
\end{figure}

The vibrational density of states (vDOS) can be calculated from the lattice Green's function used in Eq. \eqref{eq:FDT} \cite{Schirmacher2006} by using the Plemelj identity \cite{cahill2019} (see Appendix B for details): 
\begin{equation}
D(\omega)=-\frac{2\omega}{\pi}\int\text{Im}[ \langle G(\mathbf{q},\omega) \rangle ]d^dq.
\label{eq:DOS}
\end{equation} 
Taking the quenched average of the Green's function (directly associated with the DSF considered above) does not change the result of the underlying fluctuation-dissipation theorem. Combining Eq. \eqref{eq:FDT} and Eq. \eqref{eq:quenchS1}, and again benefiting from the use of $\delta$-function under the integral over disorder, we obtain the average (observable) vDOS of the longitudinal acoustic branch, which presents a characteristic feature of a peak at low frequency, in excess of the classical Debye law in the zero-disorder case:
\begin{equation}
D(\omega)\propto\omega\int \exp\left[\frac{-(\omega^2-\omega^2_{A,L}(\mathbf{q},k_0))^2}{2({\sigma}/k_0)^2 \omega^4_{A,L}(\mathbf{q},k_0)}\right]d^dq ,
\label{eq:DOSquench}
\end{equation}
where we continue to use the approximate frequency $\omega^2_{A,L} = (4k_0/M) \sin^2 (qa/2).$

The origin of this peak in our case is the Gaussian distribution of spring constant $k$, which remains in the  $D(\omega)$ even after the integration over wavevectors in Eq. \eqref{eq:DOSquench}. The position of this peak is not easy to find analytically. In Appendix, we provide the derivation of ideal vDOS, when $\sigma\rightarrow0$. Figure \ref{fig:BP} shows the vDOS predicted by Eq. \eqref{eq:DOSquench} on systems of varying strength of (Gaussian) random disorder in bond force constants. Clearly, the characteristic peak position shifts with the increasing disorder strength. The weaker the disorder (smaller ${\sigma}$), the narrower is the peak, which ultimately blows up at a resonant frequency when $\sigma \to 0$. In other words, the characteristic peak disappears when the system returns to the perfect crystalline solid. It might be tempting to associate this feature with the celebrated ``boson peak'', which is a universal feature in amorphous and glassy solids \cite{Shintani2008,Schirmacher1998} (and in some non-centrosymmetric crystals~\cite{Milkus2016,Cui2019Nonaffine} and anharmonic crystals~\cite{Baggioli2019}), since the peak lies in the region where the boson peak is usually expected. Although the traditional "boson peak" might be interpreted differently with distinct characteristics, it might be that some of the observed peaks in this range are of this origin.
Same as the model of quenched random average of DSF, Eq. \eqref{eq:DOSquench} applies to a general springs system with Gaussian distribution of spring constants.
A similar instability in lattices with disorder in spring constants has been recently observed in \cite{Mizuno2020} as a result of proliferation of negative spring constants.

\section{Conclusion}

In this article, upon explicitly calculating the quenched average of the longitudinal DSF (and also the corresponding vibrational DOS) with the representative Gaussian distribution of random bond constants in a perfect lattice, we obtained the analytical expression for the effective sound damping coefficient: $\Gamma (q) = \sqrt{2} {\sigma} \Omega(q) /k_0$, where $k_0$ is the average elastic spring constant of the bonds, and ${\sigma}$ is the variance of the random distribution of spring constants, with {$\Omega(q)$ the phonon propagation frequency.} The unexpected linear $q$-dependence of the damping coefficient $\Gamma$ was obtained, and its origin rationalized as due to the linear relation between the peak broadening and the phonon resonance frequency (at low disorder), in Eq. \eqref{eq:lineardamping}. Note that the linear scaling in Landau-Rumer theory is very different, due to quantum effects, while our theory is completely classical. This linear scaling is in contrast to the known $q^2$-dependent broadening of $S(q,\omega)$ frequently reported in glasses~\cite{Montagna1999,Baggioli} or in anharmonic materials~\cite{Baggioli2019}.  The physical system we are considering here is rather different from a typical glassy solid, because we are working with a relatively low disorder in bonds in an otherwise perfect background lattice. It is also possible that the periodic boundary conditions imposed on the lattice in many simulations (as in \cite{Montagna1999}), may affect the behavior of the collective (acoustic) modes in reciprocal space. While we focused on the formula for the longitudinal dynamical structure factor, our method is easily extended to the case of transverse DSF, where similar linear damping is expected. 

We also remark that the mechanism of phonon transport we have revealed in this paper, in both analytical and numerical ways, does not contradict with the well-known Anderson localization, even in the 1D case, where electron or light waves become spatially localized when the total number of sites in the lattices grows \cite{Anderson1958,Thouless1972}. The elastic wave considered here propagates globally, as a result of the homogeneous, cooperative motion of lattice points in a finite size system. The broadening coming from the level repulsion between “localized” modes in the random matrix theory is completely different form the wave we are looking for.

With the analytical formula for the estimated damping constant available, one can also work backwards and reveal the extent of disorder in a material using the scaling law of damping coefficient. In a continuous medium, the disorder reflects the variation of local elastic modulus $C$.  For example, in $\alpha$-quartz, the (largest) linewidth is measured around the resonance of $10$meV along $(1,0,0)$ to have the width $\Gamma\sim 5\text{meV}$ \cite{Burkel2001}. The Young's modulus $C_{11}$ was computed to be $90$GPa \cite{Cui2019Nonaffine}, therefore, this gives us an estimate for the variance $\sigma\sim 30$GPa $= 0.3C_{11}$,  a reasonable value of `low disorder'.

Very recently a damping coefficient which scales linearly with frequency in granular packings~\cite{Zhai2020}, and hence $\Gamma \sim q$ as the acoustic branch has $\omega = cq$ at low $q$. This is in full agreement with our predicted damping law in Eq. \eqref{eq:lineardamping}. Importantly, the linear damping has been observed in packings with heterogeneous distributions of contact stiffness, whereas in packings with uniform contact stiffness the linear damping is less visible. 
Considering the wave propagation with large wavevectors will be a possible future direction. 
Finally, the characteristic peak arising from random disorder in the lattice spring is due to quasi-localized vibrations directly caused by quenched disorder, and provides a possible unifying route to the boson peak phenomenon in glasses, defective crystals as well as anharmonic crystals. 

\section*{Acknowledgements}
BC acknowledges the support of the CSC-Cambridge Scholarship. AZ gratefully acknwoledged financial support from US Army Research Office (ARO) through contract nr. W911NF-19-2-0055.

\begin{appendix}
\renewcommand\thefigure{\thesection.\arabic{figure}} 
\section{Dynamical structure factor for the 1D linear chains}
\setcounter{figure}{0} 
Figure \ref{fig:opticalmodes}(a) shows the result of numerical simulation of a 1D chain with the same springs ($k_0$) connecting nearest neighbours of equal mass $M$ separated by equal distance $a$. The $q$ dependent eigenvectors/eigenvalues are calculated numerically from  the symmetric tridiagonal dynamical matrix of this simple system, before they are substituted into Eq. (2) in the main text. It is clear in Fig. \ref{fig:opticalmodes}(a) no damping takes place at any $q$. The divergence of $S(q,\omega)$ as $\omega \rightarrow 0$ is due to elastic $q=0$ scattering.

\begin{figure}
\centering
\includegraphics[width=0.45\textwidth]{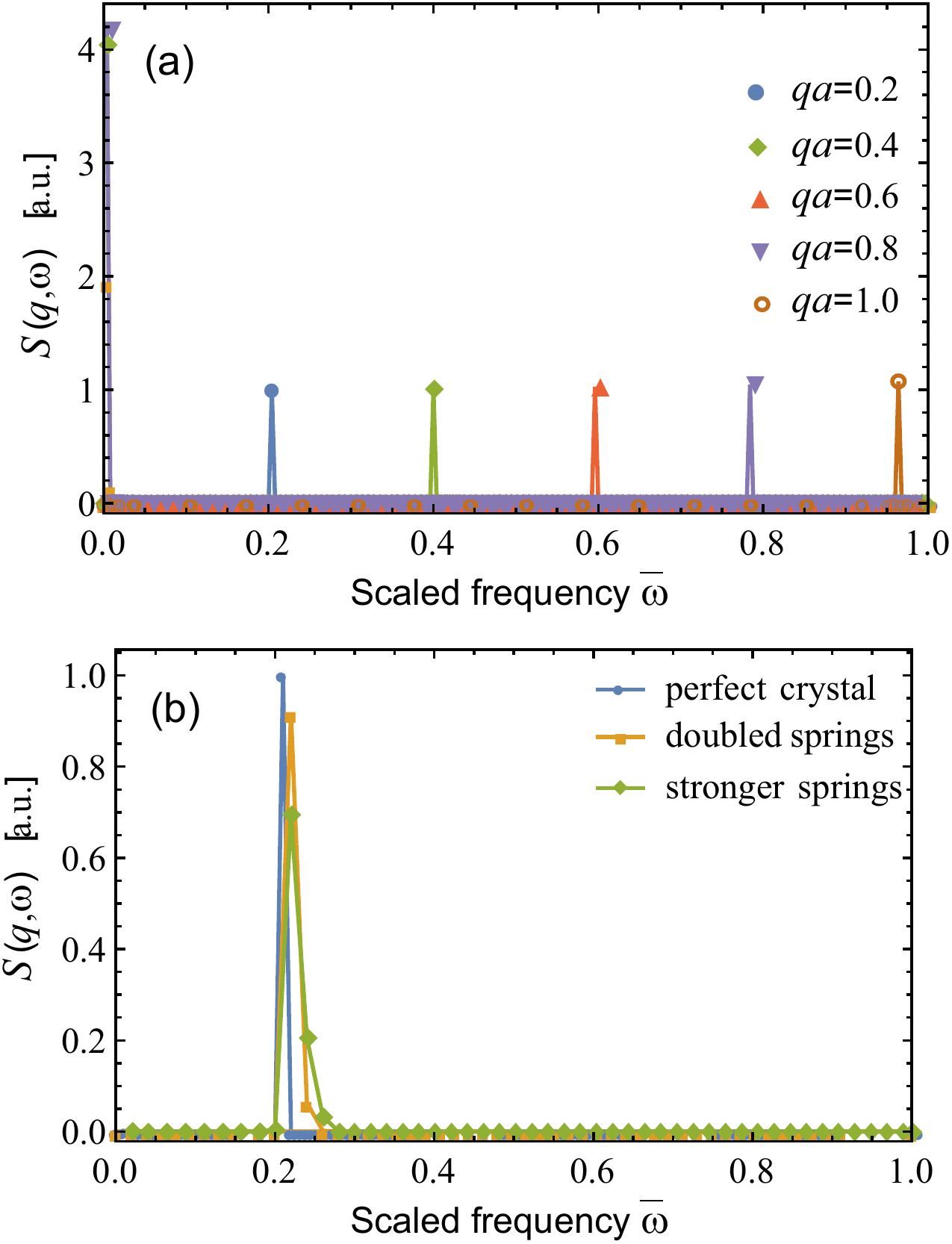}
\caption{The dynamical structure factor of 1D linear chains with and without damping, plotted against the non-dimensional frequency scaled by a factor $\sqrt{k_0/M}$. There are 3000 masses separated in equal distance in a unit cell. (a) DSF for several values of wavevector $q$, cf. Eq. (7).  (b) DSF with added disorder in some springs. A hundred springs have
their spring constants doubled, then further two hundred are replaced with three (100 out of 200) and four times (remaining 100) spring constants compared with the original one; the wavenumber $qa=0.2$.}
\label{fig:opticalmodes}
\end{figure}
 
When the interaction is harmonic, the occurrence of sound attenuation is due to defects, or generally structural disorder, which we examine below. In Fig. \ref{fig:opticalmodes}(b), we show  a preliminary indication of our main results, showing the computed DSF with a certain model disorder in springs:  $3\%-10\%$ (randomly chosen) springs were replaced with those of larger (twice or higher) spring constant. The more springs are replaced and the larger spring constants they have, the wider the curve becomes. We also find that, adding disorder does not change the position of the structural peak of DSF at the resonant frequency, which is expected given the form of $\Omega(q)$ used in Eqs.(4-5) in the main text.

\section{Derivation of Eq. (14) in the main article}
The vibrational density of states (DOS) is defined as
\begin{equation}
D(\omega)=\frac{1}{3N}\sum_\lambda\delta(\omega-\Omega_{\lambda})
\end{equation}
where he index $\lambda$ representing the pair of indices $j\textbf{q}$ with $j$ the branch and $\textbf{q}$ the wavevector. Here $N$ denotes the total number of atoms in the solid.
It is convenient to work with the distribution function of the eigenfrequencies squared since the eigenvalue distribution of the equation of motion of lattice dynamics is expressed in terms of the eigenfrequencies squared. In the harmonic case, the diagonalization of the dynamical matrix provides the bare (non-renormalized) eigenfrequencies squared as the eigenvalues of the dynamical matrix. 
Hence, we have:
\begin{equation}
\rho(\omega^{2})=\frac{1}{3N}\sum_\lambda\delta(\omega^{2}-\Omega_{\lambda}^{2})
\end{equation}
and therefore:
\begin{equation}
D(\omega)=2\omega \rho(\omega^{2}).
\end{equation}
From this we have,
\begin{equation}
D(\omega)=\frac{2\omega}{3N}\sum_\lambda\delta(\omega^{2}-\Omega_{\lambda}^{2}).
\end{equation}

We can now recall the form of the Green's function anharmonic phonons,
\begin{equation}
G_{L,T}(q,\omega)=\frac{-1}{\omega^{2}-q^{2} v_{L,T}^{2}-i\epsilon},\label{Gf}
\end{equation}
where $\epsilon$ is small,
and we also recall the well known Plemelj formula:
\begin{equation}
\frac{1}{x-x'\pm i\epsilon}=\mp i\pi\delta(x-x')+\frac{P}{x-x'},
\end{equation}
where $P$ denotes the principal part and $\frac{P}{x-x'}$ is real. 

Since $\sum_{\lambda}...\equiv\sum_{jq}$ is essentially a sum over the wavevector $q$ (we are now specializing on an isotropic crystal or solid), we can replace the discrete sum over $q$ with an integral over $q$, and upon combining Eq.(B4), Eq.(B5) and Eq. (B6) above we get:
\begin{equation}
D(\omega)=-\frac{2\,\omega}{3\,\pi
\,N}\int_0^{q_D}dq^3\,\text{Im}\left\{2\,G_T(q,\omega)+G_L(q,\omega)\right\},\label{eq8}
\end{equation}
which is Eq. (14) in the main article, and where $q_{D}$ denotes the maximum (Debye) wavenumber in the system, $q_{D}=(6\pi^2 N/V)^{1/3}$ and $N$ the number of atoms in the system.

\section{Ideal DOS}
The ideal DOS is obtained when $\sigma\rightarrow0$, which reads
\begin{align}
D(\omega)&\propto \omega \int \delta(\omega^2-\omega_{A,L}^2(q))d^dq\notag\\
&\propto\int \delta(\omega-\omega_{A,L}(q))d^dq
\label{DOS}
\end{align}
for non-negative $\omega$. The last equality can also be obtained by Eq. (7) in the main text.
We have proposed an estimate for $\omega_{A,L}$ in the main text, which is qualitatively equal to case of standard linear chains:
\begin{equation}
\omega_{A,L}=2\sqrt{\frac{k_0}{M}}|\sin\left(\frac{qa}{2}\right)|.
\end{equation}
Substitute this back to Eq. \eqref{DOS}, we can find the analytic formula of ideal DOS in 3D.
\begin{align}
&D(\omega)\propto\int_0^{q_D} q^2\delta\left(\omega-2\sqrt{\frac{k_0}{M}}|\sin\left(\frac{qa}{2}\right)|\right)dq\notag\\
&=\int_0^{|\sin\frac{q_Da}{2}|} (\frac{2}{a}\arcsin x)^2\delta\left(\omega-2\sqrt{\frac{k_0}{M}}x\right)\frac{2}{a}\frac{1}{\sqrt{1-x^2}}dx\notag\\
&=\int_0^{|\sin\frac{q_Da}{2}|} (\frac{2}{a}\arcsin x)^2\delta\left(x-\frac{\sqrt{M}}{2\sqrt{k_0}}\omega\right)\sqrt{\frac{M}{k_0}}\frac{1}{a}\frac{1}{\sqrt{1-x^2}}dx\notag\\
&=\left(\frac{2}{a}\arcsin(\frac{\omega\sqrt{M}}{2\sqrt{k_0}})\right)^2\sqrt{\frac{M}{k_0}}\frac{1}{a}\frac{1}{\sqrt{1-\frac{M}{4k_0}\omega^2}}\notag\\
&=\frac{8}{a^3}\frac{\left(\arcsin(\frac{\omega\sqrt{M}}{2\sqrt{k_0}})\right)^2}{\sqrt{\frac{4k_0}{M}-\omega^2}}.
\end{align}
In the 2nd line, we have let $x=|\sin\frac{qa}{2}|$ and require
\begin{equation}
\omega\leq2\sqrt{\frac{k_0}{M}}|\sin\left(\frac{q_Da}{2}\right)|.
\end{equation}
As $\omega\rightarrow0$, we have $D(\omega)\sim \omega^2$. The Debye law is recovered in low regime of frequency.
\end{appendix}

\bibliography{damping}

\end{document}